\documentclass[prd,twocolumn,showpacs,amsmath,amssymb]{revtex4}
\usepackage{graphicx}
\usepackage{dcolumn}
\usepackage{bm}

\begin{document}

\title{Thermodynamics in Loop Quantum Cosmology}
\author{Li-Fang Li}
  \affiliation{Department of Physics, Beijing Normal University, Beijing 100875, China}

\author{Jian-Yang Zhu}
\thanks{Author to whom correspondence should be addressed}
  \email{zhujy@bnu.edu.cn}
  \affiliation{Department of Physics, Beijing Normal University, Beijing 100875, China}

\begin{abstract}
Loop quantum cosmology (LQC) is very powerful to deal with the
behavior of early universe. And the effective loop quantum cosmology
gives a successful description of the universe in the semiclassical
region. We consider the apparent horizon of the
Friedmann-Robertson-Walker universe as a thermodynamical system and
investigate the thermodynamics of LQC in the semiclassical region.
The effective density and effective pressure in the modified
Friedmann equation from LQC not only determine the evolution of the
universe in LQC scenario but are actually also found to be the
thermodynamic quantities. This result comes from the energy
definition in cosmology (the Misner-Sharp gravitational energy) and
is consistent with thermodynamic laws. We prove that within the
framework of loop quantum cosmology, the elementary equation of
equilibrium thermodynamics is still valid.
\end{abstract}


\maketitle

\section{Introduction}

Loop quantum gravity (LQG) \cite
{ashtekar04,Smolin,rovelli98,thiemann01} is a non-perturbative and
background independent quantization of gravity. One of the
important and successful applications of LQG is loop quantum
cosmology (LQC). It has been shown that LQC resolves the problem
of classical singularities both in an isotropic model
\cite{bojowald01} and in a less symmetric homogeneous model \cite
{bojowald03}. LQC also gives a quantum suppression of classical
chaotic behavior near singularities in Bianchi-IX models \cite
{bojowald04a,bojowald04b}. Furthermore, it has been shown that
non-perturbative modification of the matter Hamiltonian leads to a
generic phase of inflation \cite{bojowald02,date05,xiong1}. On the
other hand, we know that spacetime's thermodynamic properties
result from, in a sense, quantum effects of spacetime
\cite{hawking75}. Therefore, it is very interesting and important
to investigate the thermodynamics of quantum gravity. There are
indeed many results on thermodynamical implications of loop
quantum gravity \cite {rovelli96,ghosh06}, but very little
discussion on the thermodynamics of loop quantum cosmology, which
will be the focus of the present paper.

In LQC, the phase space for spatially flat universe is spanned by
coordinates $c=\gamma \dot{a}$, being the gravitational gauge
connection, and $p=a^2$, being the densitized triad. $\gamma $ is
the Barbero-Immirzi parameter, and $a$ is the scale factor of the
universe. In the LQC scenario, the evolution of the universe can
be divided into three phases: (i) Initially, there is a truly
discrete quantum phase which is described by a difference equation
\cite{ashtekar03}. In this stage the universe may be
non-equilibrium due to the fast quantum evolution. (ii) As the
volume of universe increases and matter density decreases, the
discrete quantum effect becomes less important, the universe
enters an intermediate semi-classical phase in which the evolution
equations take a continuous form but with modifications due to
non-perturbative quantum effects \cite{bojowald02}. In this stage,
the effective loop quantum cosmology is valid and it is reasonable
to approximately treat the universe as a thermodynamic system in
equilibrium. The thermodynamic properties are subject to quantum
effects, and we are most interested in this stage. (iii) Finally,
there is the classical phase in which the quantum effects vanish
and the usual continuous equations describing cosmological
behavior are established and so is the usual thermodynamics
\cite{gibbons77,cai05}.

In recent years, many authors endeavor to study the thermodynamics
\cite{rovelli96,ghosh06} of black holes in the semiclassical
context and the framework of LQG. Now people have studied the
effective theory, though not complete yet due to quantum
back-reaction \cite{bojowald08}, in loop cosmology. Thus there is
considerable interest in the thermodynamic properties of the
universe in LQC scenario. With the universe being non-stationary
and evolving, the thermodynamics is different from the black hole
systems. It is conceivable that some of the mechanisms involved in
establishing thermal equilibrium may be modified, especially when
the expansion time scale becomes comparable to that of the matter
processes responsible for establishing the thermal equilibrium.

To resolve this issue, we develop a procedure to study the
thermodynamic properties at the apparent horizon of a
Friedmann-Robertson-Walker (FRW) universe. Our analysis is based
on the effective theory of LQC and the homogeneous and isotropic
cosmological setting. Fundamentally, comparing the modified
Friedmann equation with the ordinary one, we derive the effective
density and pressure of the perfect fluid. Then, we introduce the
Misner-Sharp energy \cite{misner64}, which is different from the
other forms of energy for its relation to the structure of the
spacetime and one can relate it to the Einstein equation. From the
expression of the Minser-Sharp energy, we get the physical meaning
of the effective density. Further more, from the conservation law,
we get the physical meaning of the effective pressure. To
understand the intrinsic essence of the effective density and
pressure, we prove that within the framework of loop quantum
cosmology, the fundamental relation of thermodynamics is still
valid.

This paper is organized as follows. In Sec. \ref{Sec.2}, we briefly
review the framework of the effective LQC. We present the dynamics
in terms of effective density and pressure, which will be defined
there. Then in Sec. \ref{Sec.3}, we obtain the thermodynamic origin
of the effective density and pressure. Some elementary consequences
are also noted. In Sec. \ref{Sec.4} we conclude this paper with some
discussions on the further implications for phenomenology.

\section{A short review of Effective theory of LQC}

\label{Sec.2}In this section, we give a short review of the
effective framework of LQC before we study the thermodynamics. The
classical form of Hamiltonian for spatially flat universe is
\begin{equation}
H_{cl}=-\frac 3{8\pi \gamma ^2}\sqrt{p}c^2+H_M\left( p,\phi
\right).
\end{equation}
There are two kinds of important modifications in the LQC. The
first one is based on the modification to the behavior of inverse
scale factor below a critical scale factor (the inverse volume
modification). The second one essentially comes from the discrete
quantum geometric nature of spacetime (quadratic modification), as
predicted by the LQG. Besides of these two kind of corrections,
there is also the more generic quantum back-reaction which gives
rise to effective potentials. In this paper, we only consider the
corrections coming from the quadratic modification. But it is
worthy to note that our result is valid for general effective
potential. With the quadratic modification, the effective
Hamiltonian becomes \cite{ashtekar06,mielczarek08}
\begin{equation}
H_{eff}=-\frac 3{8\pi \gamma ^2\bar{\mu}^2}\sqrt{p}\sin ^2\left( \bar{\mu}%
c\right) +H_M\left( p,\phi \right),
\end{equation}
The variable $\bar{\mu}$ corresponds to the dimensionless length
of the edge of the elementary loop and is given by
\begin{equation}
\bar{\mu}=\xi p^\lambda,
\end{equation}
where $\xi>0$ and $\lambda$ depend on the particular scheme in the
holonomy corrections. In this paper we take $\bar{\mu}$-scheme,
which gives
\begin{equation}
\xi^2=2\sqrt{3}\pi\gamma l_p^2
\end{equation}
 and $\lambda=-1/2$, where $l_p$ is Planck length. With this effective Hamiltonian, we
have the canonical equation
\begin{equation}
\dot{p}=\left\{ p,H_{eff}\right\} =-\frac{8\pi \gamma }3\frac{\partial
H_{eff}}{\partial c},
\end{equation}
or,
\begin{equation}
\dot{a}=\frac{\sin (\bar{\mu}c)\cos (\bar{\mu}c)}{\gamma \bar{\mu}}.
\end{equation}
We define energy density and pressure of matter \cite{hossain05}
as
\begin{eqnarray}
\rho =a^{-3}H_M ,\ P =-\frac 13a^{-2}\frac{\partial H_M }{
\partial a}.
\end{eqnarray}
Combining with the constraint on Hamiltonian, $H_{eff}=0$, we
obtain the modified Friedmann equation,
\begin{equation}
H^2=\frac{8\pi }3 \rho \left( 1-\frac \rho {\rho _c}\right)
\label{friedmann1}
\end{equation}
where $H\equiv\frac{\dot{a}}a$ denotes the Hubble rate, and $\rho
_c\equiv\frac 3{8\pi \gamma ^2\bar{\mu}^2p}$ is the quantum
critical density. Compared with the standard Friedmann equation,
we can define the effective density
\begin{equation}
\rho _{eff}=\rho \left( 1-\frac \rho {\rho _c}\right) .  \label{rhoeff}
\end{equation}
Taking derivative of Eq.(\ref{friedmann1}) and also using the
conservation equation of matter $\dot{\rho}+3H\left( \rho
+P\right) =0$, we obtain the modified Raychaudhuri equation,
\begin{equation}
\frac{\ddot{a}}a=\dot{H}+H^2=-\frac{4\pi }3\left\{ \rho (1-\frac \rho {\rho
_c})+3\left[ P(1-\frac{2\rho }{\rho _c})-\frac{\rho ^2}{\rho _c}\right]
\right\} .
\end{equation}
Compared with the standard Raychaudhuri equation, we can define the
effective pressure,
\begin{equation}
P_{eff}=P\left( 1-\frac{2\rho }{\rho _c}\right) -\frac{\rho ^2}{\rho _c}.
\label{peff}
\end{equation}
For different quantum corrections, the $\rho_{eff}$ and $P_{eff}$
may have different form. But our following statement is still
valid. In terms of the effective density and the effective
pressure, the modified Friedmann, Raychaudhuri and conservation
equation take the following forms,
\begin{equation}
H^2=\frac{8\pi }3\rho _{eff},  \label{friedmann}
\end{equation}
\begin{equation}
\frac{\ddot{a}}a=\dot{H}+H^2=-\frac{4\pi }3\left( \rho
_{eff}+3P_{eff}\right) ,  \label{raychaudhuri}
\end{equation}
\begin{equation}
\dot{\rho}_{eff}+3H\left( \rho _{eff}+P_{eff}\right) =0.  \label{conserve}
\end{equation}
Till now, $\rho _{eff}$ and $P_{eff}$ are nothing but mathematical
symbols to denote the coupling of matter and gravity. They still
lack a thermodynamic origin, as noted by the authors of
\cite{banerjee05}. In the following, we will explore their
intrinsic meaning in the thermodynamic sense and discuss some
elementary implications based on above effective framework of LQC.
But our result is more general and independent on the form of
$\rho_{eff}$ and $P_{eff}$ which may be different when considering
different quantum correction and quantum back reaction.
\section{Thermodynamics in LQC}

\label{Sec.3}Let us begin with the effective LQC description of the
universe evolution. For a spatially homogenous and isotropic
universe described by the FRW metric, the line element is
represented by
\begin{equation}
ds^2=-dt^2+a\left( t\right) ^2\left( dr^2+r^2d\Omega ^2\right) ,
\label{line element}
\end{equation}
where $a\left( t\right) $ is the scale factor of the universe, $t$
being the cosmic time, and $d\Omega ^2$ is the metric of sphere
with unit radius. Thus it is clear that all dynamical behaviors of
the universe are determined by the scale factor $a\left(
t\right)$. The metric (\ref{line element}) can be rewritten as
\begin{equation}
ds^2=h_{ab}dx^adx^b+\tilde{r}^2d\Omega
\end{equation}
where $\tilde{r}=a\left( t\right) r$ and $x^0=t$, $x^1=r$ and the two
dimensional metric $h_{ab}=diag\left( -1,a^2\right) $.

For the FRW universe, the dynamical apparent horizon \footnote{
Without the whole evolution history of the universe, one cannot
know whether there is a cosmological event horizon. However,
apparent horizon always exists in the FRW universe since it is a
local quantity of spacetime.}, defined as the sphere with
vanishing expansion \cite{cai05}, can be determined by the
relation $h^{ab}\partial _a\tilde{r}\partial _b\tilde{r}=0$ as
\begin{equation}
R_A=\frac 1H,  \label{apparent_horizon}
\end{equation}
which coincides with the Hubble horizon in this case. According to
the definition of the surface gravity
\begin{equation}
\kappa =\frac 1{2\sqrt{-h}}\partial _a\left( \sqrt{-h}h^{ab}\partial _b%
\tilde{r}\right) ,
\end{equation}
its explicit evaluation at the dynamical apparent horizon $R_A$ of
the FRW universe reads
\begin{equation}
\kappa =-\frac 1{R_A}\left( 1-\frac{\dot{R}_A}{2HR_A}\right) .
\label{surface_gravity}
\end{equation}

We now introduce the Misner-Sharp spherically symmetric
gravitational energy $E$ , or simply the MS energy, defined in
natural units by \cite{misner64}
\begin{equation}
E=\frac r2\left( 1-h^{ab}\partial _ar\partial _br\right) ,
\label{MS_energy1}
\end{equation}
which is the total energy (not only the passive energy) inside the
sphere with radius $r$. The MS energy is a pure geometric quantity
and is extensively used in the literatures about thermodynamics of
spacetime \cite{hayward98,akbar07,cai07}. Its physical meaning and
the comparison to the ADM mass and Bondi-Sachs energy have been
given in \cite{hayward96}. For spherical space-time, Brown-York
energy \cite{Brown-York} agrees with the Liu-Yau energy
\cite{Liu-Yau}, but they both differ from the MS energy. For
example, for the four-dimensional Reissner-Nordstr\"{o}m black
hole, the MS energy differs from the Brown-York or Liu-Yau mass by
a term which is the energy of the electromagnetic field inside the
sphere, as discussed in \cite{hayward96}.

In terms of the apparent horizon radius (\ref{apparent_horizon}),
the Friedmann Equation (\ref{friedmann}) can be rewritten as
\begin{equation}
\frac 1{R_A^2}=\frac{8\pi }3\rho _{eff}.  \label{eq20}
\end{equation}
Now we consider the MS energy (\ref{MS_energy1}) within the apparent horizon
$r=R_A$ of the FRW universe, given by
\begin{equation}
E=\frac{R_A}2.
\end{equation}
Using Eq.(\ref{eq20}), we get
\begin{equation}
E=\frac{4\pi R_A^3}3\rho _{eff}=\rho _{eff}V.  \label{misner_energy}
\end{equation}
It shows that it is reasonable to say that $\rho _{eff}$ is indeed
the energy density, not just a mathematical symbol. Then from the
conservation equation (\ref{conserve}), which implies energy and
momentum conservation, it is also reasonable to take $P_{eff}$ as
pressure. That is to say that the gravitational effects contribute
to energy density and pressure in the thermodynamical sense. In
the following we will find that this physical meaning is
consistent with the fundamental relation of thermodynamics, which
in turn supports this physical interpretation.

To examine the fundamental relation of thermodynamics in the setup
of LQC, we consider the apparent horizon of the FRW universe as a
thermodynamical system. An ansatz is made: assume that the apparent
horizon has an associated Hawking temperature $T$ and entropy $S$
expressed respectively as
\begin{equation}
T=\frac{\kappa }{2\pi },\ S=\frac A4, \label{T and S}
\end{equation}
where $A=4\pi R_A^2$ is the area of the apparent horizon.

Taking derivative of the energy equation (\ref{misner_energy}),
and using conservation equation (\ref{conserve}), we get
\begin{equation}
dE=4\pi R_A^2\rho _{eff}\dot{R}_Adt-4\pi R_A^3H\left( \rho
_{eff}+P_{eff}\right) dt. \label{dE}
\end{equation}
Beside this, by taking derivative of the Friedmann Equation
(\ref{eq20}) and using the conservation equation (\ref{conserve}),
we get the differential form of the Friedmann equation
\begin{equation}
\frac 1{R_A^3}dR_A=4\pi (\rho _{eff}+P_{eff})Hdt.
\end{equation}
Considering the surface gravity on the apparent horizon (\ref
{surface_gravity}), we can multiply both sides of the above
equation by a factor $R_A\left( 1-\frac{\dot{R}_A}{2HR_A}\right) $
and get
\begin{equation}
\frac \kappa {2\pi }d\left( \pi R_A^2\right) =-4\pi R_A^3\left( \rho
_{eff}+P_{eff}\right) H\left( 1-\frac{\dot{R}_A}{2HR_A}\right) dt.
\label{kdT}
\end{equation}
Therefore, in virtue of the ansatz (\ref{T and S}), and combining
the Eqs. (\ref{dE}) and (\ref{kdT}), one gets
\begin{equation}
dE=TdS+WdV,
\end{equation}
where $W=\frac{\rho_{eff}-P _{eff}}2$ is the work density if we take $%
\rho _{eff}$ and $P_{eff}$ as the energy density and pressure
physically \cite{hayward98}. Again, we see that taking $\rho _{eff}$
and $P_{eff}$ as the energy density and pressure in the
thermodynamical sense is consistent with
the fundamental relation of thermodynamics. However, if we take $\rho $ and $%
P $ as the thermodynamical quantities, we will find that
\begin{equation}
dE=TdS+W^{\prime }dV+\frac{\rho}{\rho _c}PdV,
\end{equation}
with work density $W^{\prime }=\frac{\rho-P}2$. This equation
means that the fundamental relation of thermodynamics breaks down
unless we consider the work term now does not take the form
suggested by \cite{hayward98}. But this complicated expression for
work term seems not reasonable. In contrast, the physical
interpretation that $\rho _{eff}$ and $P_{eff}$ are
thermodynamical quantities in LQC is consistent with the
fundamental relation of thermodynamics. Or to say in terms of
thermodynamical quantities $\rho _{eff}$ and $P_{eff}$, the
fundamental relation of thermodynamics is valid in LQC too.

\section{Conclusion}

\label{Sec.4}In Conclusion, we have investigated the thermodynamic
properties of the universe in LQC scenario and found that the
fundamental relation of thermodynamics is valid in the effective LQC
scenario. We found that the effective density $\rho _{eff}$ and the
effective pressure $P_{eff}$ are not only a symbol to denote the
coupling between the gravity and matter, but actually the energy
density and pressure in thermodynamical sense. This result comes
from the energy definition in cosmology (the Misner-Sharp
spherically symmetric gravitational energy) and is consistent with
fundamental relation of thermodynamics.

In the following, we briefly comment on the physical meanings from
the expressions of the effective energy density and pressure. When
the energy density is much smaller than the quantum critical
density ($\rho \ll \rho _c$), the effective density $\rho _{eff}$
and the effective pressure $P_{eff}$ come back to the traditional
ones, i.e. $\rho$ and $P$, and the classical picture is recovered.
Apart from the contribution of the matter sector, the effective
density and pressure also receive the contribution from the
spatial curvature. Also note that while for large volumes the
spatial curvature is negligible to the density and pressure, for
small volumes it is important. Since the $\rho _{eff}$ and
$P_{eff}$ have thermodynamic meanings, and non-perturbative
modification to the matter field at short scales implies inflation
which also means a violation of the strong energy condition
\cite{xiong2}, we can expect that the wormhole solution maybe a
normal object in effective LQC. Similarly, the spectrum of
fluctuation of $\rho _{eff}$ may be more important than $\rho $
itself which contributes to the large-scale structure of the
universe. All these are interesting topics for further study.

\acknowledgements The work was supported by the National Natural
Science Foundation of China (No.10875012). L Li is indebted to Dr.
Dah-Wei Chiou for his helpful discussions.

\end{document}